\pdfoutput=1
\documentclass[journal]{IEEEtran}

\usepackage{graphicx}
\usepackage[caption=false,font=footnotesize]{subfig}
\usepackage{tabularx}
\usepackage{array}
\usepackage{amsmath,amssymb}
\usepackage{cite}
\usepackage{url}
\usepackage{balance}
\usepackage{microtype}

\title{Movable-Element STAR-RIS for Integrated Sensing and Communication: Architectures, Opportunities, and Practical Challenges}
\author{Wali Ullah Khan,~\IEEEmembership{Member,~IEEE} and Muhammad Adil 
\thanks{Wali Ullah Khan is with SnT, University of Luxembourg, Luxembourg (e-mail: waliullahkhan30@gmail.com).}
\thanks{Muhammad Adil is with the Department of Electronics Engineering, University of Rome Tor Vergata, 00133 Rome, Italy (e-mail: adil34700@gmail.com).}
}

\begin{document}
\maketitle

\begin{abstract}
Simultaneously transmitting and reflecting reconfigurable intelligent surfaces (STAR-RISs) extend conventional reflecting-only surfaces by enabling controllable full-space propagation. Yet, once deployed, the physical locations of their elements remain fixed, leaving the surface geometry unable to adapt to users, sensing targets, blockage, or near-field focusing conditions. This article develops a system-level perspective on movable-element STAR-RIS (ME--STAR--RIS) for integrated sensing and communication (ISAC), where the surface jointly reconfigures its electromagnetic response and the physical positions of its elements. We explain how geometric reconfiguration can reshape the effective aperture, spatial correlation, interference nulls, and sensing illumination while preserving STAR-RIS full-space operation. A two-timescale control architecture separates relatively slow element motion from fast beamforming and transmission/reflection control. A 100-realization illustrative case study compares ME--STAR--RIS with an otherwise identical fixed STAR-RIS under a passive coupled transmission/reflection response. The results show a substantially improved sampled communication--sensing tradeoff and, importantly, a rapid saturation of the rate gain with modest element travel. We conclude with representative use cases, implementation constraints, and a research roadmap covering mobility overhead, channel acquisition, mutual coupling, near-field operation, hardware impairments, and learning-assisted predictive control.
\end{abstract}


\section{Introduction}
Integrated sensing and communication (ISAC) is widely regarded as a central component of future wireless networks because it allows communication and radio sensing to share spectrum, hardware, waveform resources, and network infrastructure \cite{liu2022isac}. In parallel, reconfigurable intelligent surfaces (RISs) have created a new design dimension by making the propagation environment partially programmable rather than treating it as an immutable channel \cite{wu2019irs}. These two directions are naturally complementary: ISAC demands controllable illumination, coverage, interference, and spatial resolution, while RIS technology provides a way to shape the radio environment without deploying a complete additional radio-frequency chain.

Conventional RISs, however, primarily manipulate the reflected field and therefore favor users and targets located on one side of the surface. Simultaneously transmitting and reflecting RISs (STAR-RISs) address this limitation by controlling transmitted and reflected fields and thereby providing full-space service \cite{liu2021star360}. Practical STAR-RIS models also reveal that transmission and reflection coefficients can be physically coupled, which must be considered when translating algorithmic gains into realizable hardware \cite{liu2022coupled}. Early studies have already demonstrated the potential of STAR-RIS for ISAC, including joint sensing/communication beamforming \cite{liu2023starisac}, waveform and coefficient co-design \cite{zhang2024joint}, high-mobility sensing and communication \cite{li2024highmobility}, and uplink full-space ISAC \cite{khan2025uplink}. A remaining limitation is less visible but equally fundamental: the element geometry is fixed after installation. Even if every STAR element can rapidly change its complex transmission and reflection response, the surface aperture, element spacing, sampling pattern, and location-dependent channel phases are predetermined by fabrication and mounting. This can be restrictive in ISAC because communication users and sensing targets can favor different spatial signatures, different focusing conditions, and different interference-null locations.

Movable antenna (MA) and fluid antenna concepts have recently shown that controlled position reconfiguration can exploit spatial channel variations as an additional degree of freedom \cite{ning2025architectures,wu2026fas}. Their relevance to ISAC has also been demonstrated in far-field and near-field settings \cite{lyu2025maisac,ding2026nearfield,chen2025position}. A related development is the fluid RIS, where the positions of RIS elements themselves become adjustable \cite{ye2025fris}. Most importantly for the topic of this article, a recent study has already considered a STAR-RIS with movable elements for integrated sensing and covert communication \cite{zhou2025mestar}. Therefore, the opportunity is no longer to claim that movable STAR elements have never been studied; rather, it is to develop a broader ISAC architecture and design perspective that explains when geometry reconfiguration is useful, how to control it, what practical constraints it introduces, and how to benchmark its gains.

Motivated by these gaps, this article develops a system-level framework for ME--STAR--RIS-enabled ISAC that jointly considers electromagnetic and geometric reconfigurability. The key motivation is that conventional STAR-RIS optimization adapts the transmission/reflection response over a fixed spatial aperture, whereas the performance of ISAC also depends on element geometry through propagation phases, spatial correlation, effective aperture, interference structure, and target illumination. This motivates treating element positions as an additional, but slower, controllable degree of freedom. In contrast to application-specific optimization studies, our objective is to clarify how geometric reconfiguration complements STAR-RIS coefficient control, identify the propagation regimes in which element mobility is most beneficial, and establish a hardware-aware framework for evaluating its communication--sensing benefits.

Accordingly, this article develops a unified perspective on ME--STAR--RIS-enabled ISAC spanning architecture, operation, control, deployment opportunities, and performance evaluation. First, we introduce a three-layer reconfigurability framework that distinguishes wavefront control, STAR transmission/reflection functionality, and element-level geometric adaptation, and position ME--STAR--RIS relative to conventional RIS, STAR-RIS, movable-antenna-assisted RIS, and movable/fluid RIS architectures. Second, we describe a full-space ISAC architecture and explain how position-dependent propagation can reshape multiuser interference, target illumination, spatial correlation, effective aperture, and near-field focusing. Third, we develop a two-timescale design and control framework in which geometry is adapted from slowly varying spatial, blockage, and mobility information, while BS beamforming and STAR coefficients track short-term channel and sensing states. Fourth, we identify representative application opportunities in vehicular, indoor, aerial, near-field, and non-terrestrial networks. Fifth, a 100-realization numerical case study under coupled STAR hardware compares ME--STAR--RIS with an otherwise identical fixed STAR-RIS and quantifies the communication--sensing tradeoff and the value of available element travel. Finally, we discuss implementation constraints and open research directions needed to translate geometric reconfigurability into practical ISAC gains.

The remainder of the article is organized as follows. Section II introduces the three layers of ME--STAR--RIS reconfigurability and positions the architecture relative to neighboring concepts. Section III describes full-space ISAC operation, hardware constraints, and position-dependent propagation, while Section IV explains the communication and sensing benefits enabled by geometric adaptation. Section V presents the two-timescale design and control framework. Section VI discusses representative application opportunities, and Section VII quantifies the mobility gain through an illustrative numerical case study. Section VIII examines practical implementation challenges and future research directions, followed by the conclusion in Section IX.

\begin{figure}[t]
\centering
\includegraphics[width=\columnwidth]{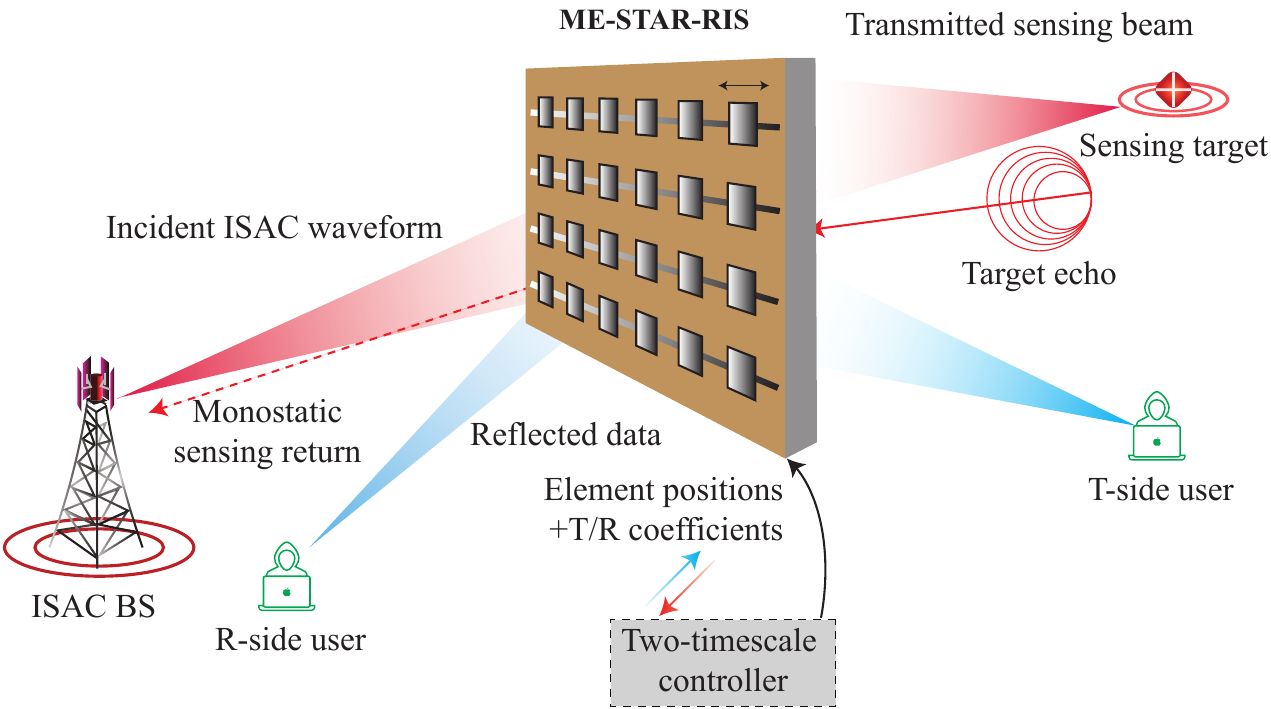}
\caption{Conceptual ME--STAR--RIS-enabled ISAC architecture with monostatic sensing at the BS. The STAR elements can move inside prescribed regions or tracks while controlling the transmitted and reflected wavefronts. For the illustrated transmission-side target, the dual-functional waveform illuminates the target through the surface and the backscattered echo returns through the surface to the same BS. A practical controller coordinates slow geometry updates with fast beamforming and transmission/reflection control.}
\label{fig:architecture}
\end{figure}

\section{From Programmable Waves to Programmable Geometry}
The distinguishing feature of ME--STAR--RIS is that reconfigurability is no longer confined to the electromagnetic response of a fixed surface. By jointly controlling the transmitted/reflected fields and the physical element geometry, the surface acquires complementary degrees of freedom that operate across the wave, functional, and spatial domains. This section explains these layers of reconfigurability and positions ME--STAR--RIS relative to closely related movable-antenna and movable-RIS architectures.

\subsection{Three layers of reconfigurability}
\begin{figure*}[t]
\centering
\includegraphics[width=\textwidth]{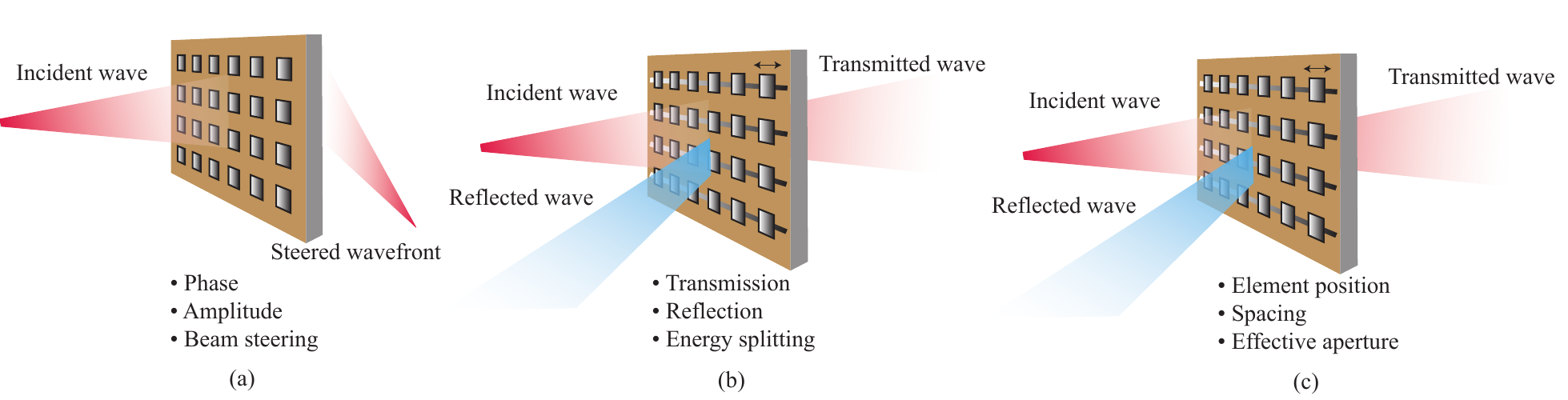}
\caption{Three layers of ME--STAR--RIS reconfigurability: (a) wavefront reconfiguration through amplitude/phase control, (b) STAR functional reconfiguration through transmission/reflection control, and (c) geometric reconfiguration through movable element positions. Their joint exploitation enables programmable propagation, full-space operation, and adaptive aperture geometry for ISAC.}
\label{fig:layers}
\end{figure*}

The evolution from RIS to ME--STAR--RIS can be understood through three increasingly rich forms of reconfigurability, illustrated in Fig.~\ref{fig:layers}.

\emph{Wavefront reconfiguration:} At the first layer, the surface changes the amplitude and phase of the reradiated field. This is the classic RIS function: steer energy toward a desired receiver or target, suppress interference in another direction, or synthesize a prescribed spatial response.

\emph{STAR functional reconfiguration:} At the second layer, the surface also controls whether and how incident energy is transmitted and reflected. Energy splitting, mode switching, and time switching can be used to serve nodes on both sides of the surface \cite{liu2021star360}. Under practical passive hardware, the transmission and reflection responses may obey coupling constraints \cite{liu2022coupled}; hence, the surface cannot be viewed as two independent colocated RISs.

\emph{Geometric reconfiguration:} The third layer introduces position control. Each element can move within a prescribed region or track while retaining its transmission and reflection controls. The electronic coefficients reconfigure the local electromagnetic response, whereas physical relocation changes propagation distances, spatial phase progression, and, in near-field or multipath conditions, even the path amplitudes seen by the incident and reradiated channels. Geometry therefore changes the channel state on which the fast electronic control subsequently operates.

This distinction is especially important in ISAC. Communication beamforming often seeks coherent signal addition and interference suppression at a small number of user locations. Sensing may seek high illumination at target directions, low sidelobes, angular/range discrimination, or a favorable Fisher-information structure. With a fixed surface, these goals must share one immutable aperture. With movable elements, the aperture becomes an additional resource.

\subsection{How is ME--STAR--RIS different from neighboring concepts?}
The key distinction is where spatial mobility is introduced and which links it changes. In movable-antenna-assisted RIS or STAR-RIS systems, the active antenna at the BS or user changes position while the surface geometry remains fixed \cite{ning2025architectures,wu2026fas}. Such mobility selects a favorable local channel at the transceiver, but it does not reshape the RIS sampling pattern or physical aperture. In ME--STAR--RIS, by contrast, moving a surface element simultaneously changes its incident BS--surface path and its outgoing paths toward users and sensing targets, so one position update can modify several cascaded links.

Movable/fluid RIS architectures also introduce programmable geometry, but are typically developed for reflecting operation \cite{ye2025fris}. ME--STAR--RIS combines that geometric freedom with simultaneous transmission and reflection. The same movable aperture must therefore support two half-spaces while respecting STAR energy-splitting and, for practical passive hardware, transmission/reflection phase-coupling constraints. A position that favors a transmission-side user or target can consequently alter reflection-side channels at the same time.

This difference is particularly relevant to ISAC, where communication users and sensing targets may lie on opposite sides of the surface or impose competing spatial requirements. ME--STAR--RIS uses fast electronic coefficients to shape the transmitted/reflected fields and slower position control to reshape the underlying aperture and channel geometry. Table~\ref{tab:comparison} summarizes this positioning relative to neighboring architectures.

\begin{table}[!t]
\caption{Positioning Relative to Neighboring Architectures}
\label{tab:comparison}
\centering
\footnotesize
\renewcommand{\arraystretch}{1.14}
\begin{tabularx}{\columnwidth}{|p{1.55cm}|p{2.25cm}|X|}
\hline
\textbf{Architecture} & \textbf{Geometry / coverage} & \textbf{Key ISAC distinction} \\ \hline
\hline
Conventional RIS & Fixed; reflection side & Programmable wavefront with mainly one-sided coverage \\ \hline
STAR-RIS & Fixed; full-space T/R & Full-space sensing/communication with immutable aperture \\ \hline
MA-assisted RIS/STAR-RIS & Surface fixed; antenna movable & Spatial channel selection at the transceiver \\ \hline
Movable/fluid RIS & Element geometry reconfigurable; typically reflective & Adaptive reflecting aperture and spatial correlation \\ \hline
\textbf{ME--STAR--RIS} & \textbf{Element geometry reconfigurable; full-space T/R} & \textbf{Programmable wavefront, STAR function, and surface geometry} \\
\hline
\end{tabularx}
\end{table}

\section{ME--STAR--RIS-Enabled ISAC Architecture}
Building on these reconfigurability dimensions, we next consider how ME--STAR--RIS can be integrated into a full-space ISAC system. The architecture must jointly account for STAR transmission/reflection operation, communication and sensing links on both sides of the surface, practical hardware coupling, and the fact that changing an element position simultaneously modifies several cascaded propagation paths.

\subsection{Full-space ISAC operation and hardware constraints}
Consider a multi-antenna ISAC base station (BS) serving communication users and sensing one or more targets with the assistance of an ME--STAR--RIS, as shown in Fig.~\ref{fig:architecture}. We adopt monostatic sensing at the BS so that communication and sensing share the same active platform. In the illustrated case, the sensing target lies on the transmission side: the BS sends a dual-functional waveform through the STAR surface toward the target, and the backscattered echo traverses the surface again on its return to the BS. Reflection-side targets are handled analogously through the reflected path. Communication users and sensing targets can therefore occupy either side of the surface while sharing the same waveform and infrastructure \cite{liu2022isac}.

The ME--STAR--RIS contains a set of controllable elements. Each element is assigned (i) a feasible movement region or track, (ii) transmission and reflection amplitude/phase controls, and (iii) a minimum separation rule that prevents element overlap and limits excessive mutual coupling. Depending on the hardware, movement may be truly mechanical, fluidic, micro-electromechanical, pixel-based/electronic-equivalent, or implemented at a subarray level rather than at every meta-atom. The implementation literature on movable and fluid antennas provides useful candidate mechanisms and reminds us that ``movability'' should not be interpreted as requiring macroscopic robotic motion in every design \cite{ning2025architectures,wu2026fas}.

For an energy-splitting STAR architecture, the incident field at an element is partitioned between transmitted and reflected components. To retain hardware realism, we focus on passive STAR implementations with energy conservation and coupled transmission/reflection responses; the numerical case study subsequently adopts one feasible coupled-phase branch. Energy conservation and practical phase coupling therefore remain active when positions are optimized \cite{liu2022coupled}. An ideal independent-phase model may be used only as an optional upper-bound reference, not as the main benchmark. This choice prevents geometry gains from being inflated by an unrealistically flexible electromagnetic model.

Full-space operation is an architectural advantage that distinguishes ME--STAR--RIS from reflecting-only movable surfaces. A sensing target may lie in the transmission half-space while a communication user lies in the reflection half-space, or vice versa. The STAR coefficients allocate and shape energy across the two regions, while the movable geometry changes the spatial response that both transmitted and reflected fields experience. The resulting design must therefore account for the fact that one physical aperture simultaneously supports communication and sensing objectives on both sides of the surface.

\subsection{Position-dependent propagation}
Moving a STAR element alters more than a steering-vector entry. In a cascaded BS--surface--user or BS--surface--target path, the element position affects the incident path from the BS and the outgoing path toward the user/target. Consequently, the same position update changes multiple cascaded channels at once. This coupling is one reason why position optimization is harder for movable RISs than for a single movable antenna. The fluid-RIS ISAC literature explicitly highlights the position dependence of both incident and reflective channels \cite{ye2025fris}.

In far-field line-of-sight conditions, the main effect of a small movement is a controllable phase perturbation determined jointly by the incident and outgoing directions. In multipath channels, movement can additionally select spatial fading peaks and alter channel correlation. In the radiative near field, position changes affect both phase and distance-dependent amplitude; geometric control can therefore modify focusing depth, focal location, and the effective Fresnel aperture. Recent near-field MA-ISAC results show that movement cost and the size of the moving region become important practical variables in this regime \cite{ding2026nearfield}. These propagation effects explain why element position is not simply interchangeable with an electronic phase shift in the regimes of greatest interest to ISAC.

\section{What Does Element Mobility Bring to ISAC?}
The benefit of ME--STAR--RIS arises from how geometric reconfiguration changes the spatial channel seen by both communication and sensing functions.

\subsection{Communication gain through channel shaping}
With fixed element positions, the STAR coefficients compensate channel phases but cannot change the spatial sampling pattern of the surface. Element motion can reduce destructive multipath combinations, reshape inter-user correlation, and provide an additional mechanism for interference suppression beyond phase-only nulling. The basic intuition is analogous to MA systems, which exploit local spatial variations to improve signal power and interference management \cite{ning2025architectures}.

An important qualification is that mobility is not guaranteed to provide a large gain in every channel. In a single-user far-field line-of-sight link with ideal continuous and independently adjustable phases, a fixed surface can already align the dominant path very effectively, leaving little extra rate gain for position control. The case for ME--STAR--RIS becomes stronger when several users or sensing targets create conflicting spatial requirements, when passive transmission/reflection phases are coupled, when multipath or spatial correlation is significant, when the surface operates in the near field, or when hardware control is quantized. These are precisely the regimes in which geometry can change more than a phase that the electronic coefficients could otherwise compensate. Wavelength-scale movement at mmWave/THz frequencies is also physically compact, making such adaptation practically interesting.

\subsection{Sensing gain through adaptive aperture geometry}
Sensing performance depends strongly on aperture geometry. A larger effective aperture generally improves angular discrimination, while nonuniform sampling can be used to control sidelobes or avoid ambiguous spatial patterns. ME--STAR--RIS can adapt the positions of a subset or all elements so that the virtual aperture is better matched to the current sensing sector. For a target close to the surface, the same degree of freedom can improve near-field focusing by modifying the spatial distribution of propagation distances.

The MA-ISAC literature already demonstrates that optimized antenna positions can improve sensing/communication tradeoffs and angle-estimation performance \cite{lyu2025maisac,chen2025position}. ME--STAR--RIS transfers this principle to a passive/semi-passive intelligent surface while simultaneously retaining full-space transmission and reflection.

\subsection{Geometry-assisted communication--sensing tradeoff}
In conventional ISAC, improving sensing may require directing more power toward a target or shaping a waveform that is less favorable for communication users. With fixed STAR geometry, transmission/reflection coefficients and BS beamformers absorb most of this conflict. ME geometry introduces a new resource: the physical surface can be repositioned to create a channel configuration in which the same fast-domain coefficients satisfy both objectives more efficiently.

This is precisely why the most informative numerical result for a magazine article is not only ``rate versus power.'' A more revealing plot is a communication--sensing tradeoff obtained by changing the relative emphasis placed on target illumination while keeping the hardware and power budgets identical for ME--STAR--RIS and fixed STAR-RIS. A favorable shift of the sampled tradeoff directly illustrates whether geometry control can deliver more communication performance at a comparable illumination level, without claiming a globally optimal Pareto frontier.

\section{Two-Timescale Design and Control Framework}
\begin{figure*}[!t]
\centering
\includegraphics[width=0.80\textwidth]{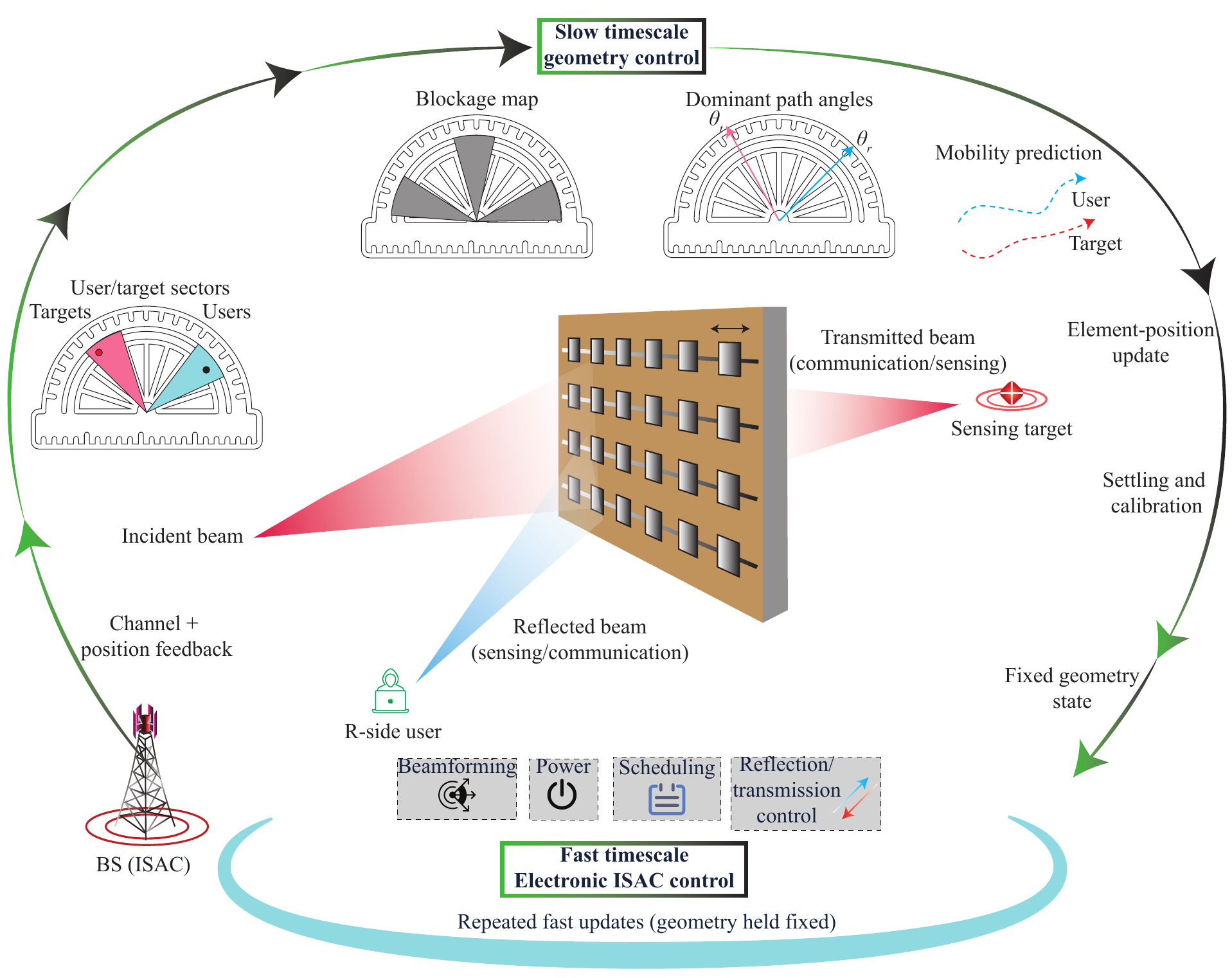}
\caption{Two-timescale design and control framework. The slow layer updates the ME--STAR--RIS geometry using user/target sectors, blockage information, dominant path angles, mobility prediction, and movement constraints. After settling and calibration, the geometry is held fixed while the fast layer repeatedly adapts beamforming, power/scheduling, and STAR transmission/reflection control to short-term communication and sensing states.}
\label{fig:timescale}
\end{figure*}

A practical ME--STAR--RIS should not reposition elements at the rate of every channel symbol. Physical motion is slower and incurs actuation, settling, and calibration overhead, whereas BS beamforming and electronic STAR control can be updated much faster. This motivates the two-timescale design in Fig.~\ref{fig:timescale}: a slow layer selects a favorable geometry state, while a fast layer repeatedly adapts the electronic ISAC variables with that geometry held fixed.

\subsection{Slow-timescale geometry adaptation}
The slow controller determines when and where the movable elements, or movable subarrays, should be repositioned. Rather than tracking instantaneous small-scale fading, it can exploit persistent spatial information such as user/target sectors, dominant angles, blockage maps, long-term channel statistics, near-field focal regions, and predicted mobility. The geometry can then be selected to strengthen desired cascaded links, reduce inter-user correlation, improve target illumination or angular discrimination, or balance communication and sensing through a joint utility.

These decisions remain constrained by the hardware: feasible positions must respect movement regions, minimum spacing, actuation energy, settling time, and positioning accuracy. Geometry updates can therefore be periodic or event driven. For example, repositioning may be triggered when a dominant path becomes blocked, a user or target enters a new angular sector, or the ISAC utility falls below a threshold. Discrete position grids, group/subarray movement, and sparse movable elements can further reduce search and actuation overhead.

\subsection{Fast-timescale electronic ISAC control}
Once the geometry has settled, the fast controller adapts BS communication/sensing beamformers, power allocation, user scheduling, and STAR transmission/reflection coefficients over short-term channel intervals. The corresponding energy-splitting, mode-switching, or time-switching constraints---including practical amplitude and phase coupling---are enforced at this layer \cite{liu2022coupled}. Existing STAR-RIS ISAC methods can thus operate within each geometry state rather than invoking physical movement whenever instantaneous CSI changes \cite{liu2023starisac,zhang2024joint}. In this hierarchy, geometry adaptation modifies the channel environment, while electronic control exploits the current channel state.

\subsection{Cross-timescale coordination and complexity}
The two layers remain coupled: a communication-favorable geometry may weaken target illumination, while a sensing-oriented aperture may increase user-channel correlation. The fast layer therefore feeds back channel quality, interference, sensing performance, and user/target state; the slow controller decides whether the expected gain justifies another movement and calibration cycle. This interaction is represented explicitly in Fig.~\ref{fig:timescale}.

From a computational viewpoint, element mobility adds position variables and movement/spacing constraints to the already nonconvex beamforming and STAR-coefficient design. Alternating optimization, projected-gradient methods, successive convex approximation, or learning-assisted search are therefore natural slow-timescale tools, while simpler fast electronic updates can be repeated more frequently. For large surfaces, position codebooks or subarray motion can further reduce complexity. Importantly, optimizing geometry for a quasi-static simulation realization quantifies the potential geometry gain; it does not imply symbol-level motion. In practice, one geometry state would be retained across many fast updates and changed only when the spatial environment evolves sufficiently.

\section{Application and Deployment Opportunities for ME--STAR--RIS-Enabled ISAC}
The potential benefits of ME--STAR--RIS are expected to be most pronounced in scenarios where the spatial environment changes over time, users and sensing targets occupy both sides of the surface, or fixed array geometry limits simultaneous communication and sensing performance. In such settings, slow geometric adaptation can complement fast electronic STAR control by reshaping the effective aperture, dominant propagation paths, and spatial coverage according to the prevailing deployment conditions. The following representative scenarios illustrate where this additional degree of freedom may be particularly valuable.

\subsection{Vehicular and roadside ISAC}
A roadside or building-mounted ME--STAR--RIS can adapt its geometry to traffic flow and target sectors over a slow timescale while rapidly controlling STAR coefficients as vehicles move. Full-space operation is attractive when some users are behind the surface and others are in front of it. A vehicle-mounted STAR-RIS has already been investigated for high-mobility ISAC \cite{li2024highmobility}; introducing element mobility could further tailor the aperture to dominant angles or compensate long-term changes in vehicle geometry and blockage.

\subsection{Indoor smart environments}
Indoor ISAC combines connectivity, localization, occupancy sensing, gesture recognition, and environment mapping. Walls and partitions naturally create users/targets on both sides of a surface. Because indoor layouts change on a slower timescale than wireless fading, they fit the proposed control hierarchy: geometry can adapt to room usage or dominant zones, while electronic STAR control follows the fast channel.

\begin{table}[t]
\caption{Illustrative Case-Study Setup}
\label{tab:sim}
\centering
\footnotesize
\renewcommand{\arraystretch}{1.05}
\begin{tabularx}{\columnwidth}{|p{2.15cm}|X|}
\hline
\textbf{Parameter} & \textbf{Value / sweep} \\
\hline\hline
Carrier / bandwidth & 28 GHz / 100 MHz \\ \hline
BS / STAR size & 8 BS antennas; 32, 64, 96, or 128 STAR elements \\ \hline
Geometry & BS $(0,0)$ m; STAR $(4,0)$ m; R-user $(2,-2)$ m; T-user $(6,-2)$ m; target $(7,2)$ m \\ \hline
STAR hardware & Equal energy splitting; single coupled T/R branch with phase difference $+\pi/2$ \\ \hline
Propagation & Rician: BS--STAR $K=10$ dB, STAR--user $K=5$ dB; LoS target; blocked direct paths; 61.4-dB loss at 1 m, exponent 2.2 \\ \hline
Noise & $-174$ dBm/Hz PSD; 7-dB noise figure \\ \hline
BS beamforming & Two ZF data beams + user-null-space sensing beam; 80\%/20\% power split \\ \hline
Mobility & Out-of-plane normal tracks; available track length $D=0$--$1\lambda$; displacement bounded by $\pm D/2$; tangential spacing $\lambda/2$ \\ \hline
Fair baseline & Same grid, elements, power, hardware, channels, solver, and 320-update phase budget \\ \hline
Protocol / metrics & 100 channel realizations; sum rate, normalized target illumination power, sampled tradeoff, and normalized displacement \\
\hline
\end{tabularx}
\end{table}

\subsection{UAV and aerial networks}
Aerial links can experience strong LoS components and rapidly changing angles. Even modest geometric changes may therefore create predictable phase shifts that complement STAR coefficient control. Lightweight subarray movement or electronically emulated position reconfiguration may be more practical than independent mechanical motion for every element. The key design question is whether the geometry update timescale can track changes in the dominant angular structure without excessive movement energy.

\subsection{Near-field and extremely large surfaces}
As apertures grow and carrier frequencies increase, users and targets can lie in the radiative near field. In this regime, a fixed regular grid may be far from optimal for simultaneous focusing at multiple communication and sensing points. Movable-element geometry can adapt the spatial sampling density and path-length distribution. Near-field MA-ISAC research indicates that movement-region size, movement matching, and cost become central design variables \cite{ding2026nearfield}; the same issues should be studied for large ME--STAR--RIS apertures.

\subsection{Non-terrestrial and integrated networks}
Satellite, HAPS, UAV, and terrestrial layers produce large variations in incident angles and service geometry. A reconfigurable STAR surface on a building, platform, or vehicle could use slow element/subarray geometry updates to track persistent spatial changes while the electronic coefficients respond to fast channel variation. For NTN applications, robust positioning, vibration tolerance, calibration, and actuation reliability may matter more than maximizing movement range.

These scenarios illustrate that the value of ME--STAR--RIS is strongly dependent on propagation geometry, mobility, blockage, and the competing spatial requirements of communication and sensing. The next section therefore moves from qualitative opportunities to a controlled numerical comparison that isolates the value of geometric reconfiguration against an otherwise identical fixed STAR-RIS.

\begin{figure*}[t]
\centering
\subfloat[Rate vs. transmit power.\label{fig:rate_power}]{\includegraphics[width=0.47\textwidth]{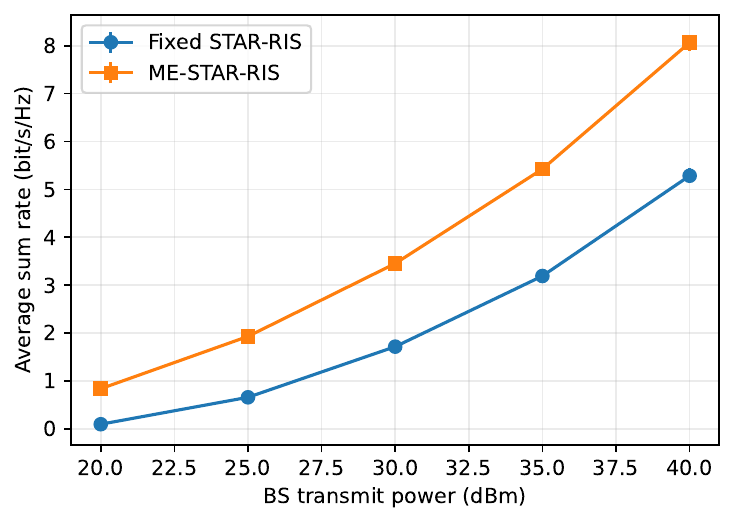}}\hfill
\subfloat[Normalized target illumination power vs. surface size.\label{fig:illum_n}]{\includegraphics[width=0.47\textwidth]{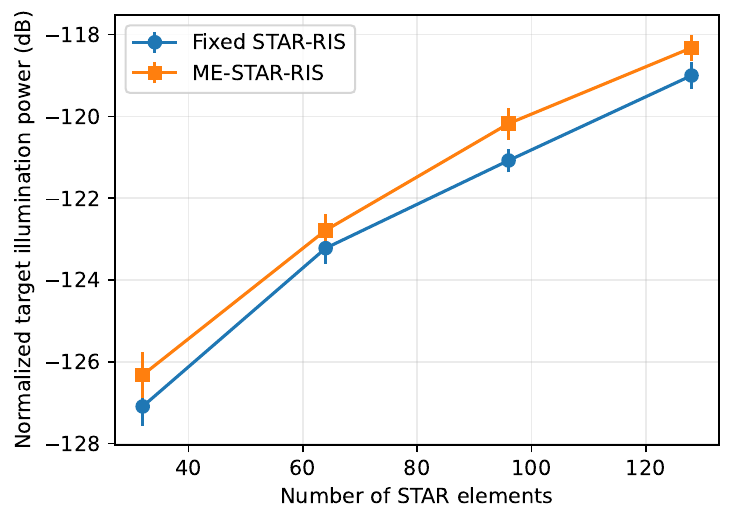}}\\[-0.5ex]
\subfloat[Communication--sensing tradeoff.\label{fig:tradeoff}]{\includegraphics[width=0.47\textwidth]{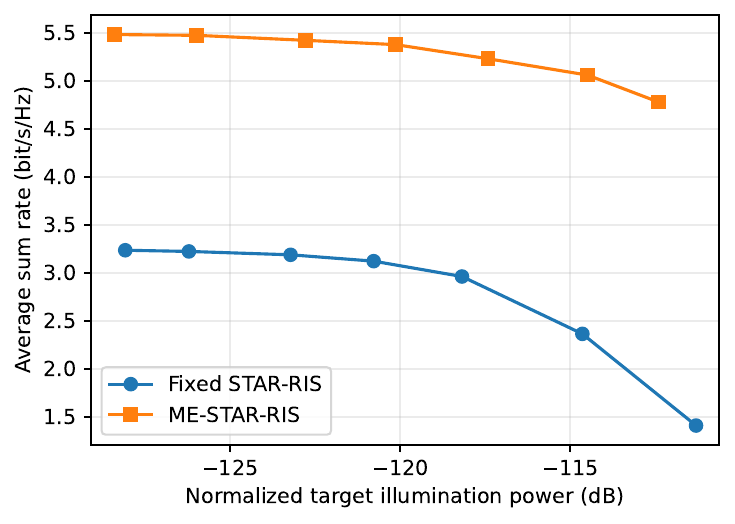}}\hfill
\subfloat[Rate vs. available track length.\label{fig:track}]{\includegraphics[width=0.47\textwidth]{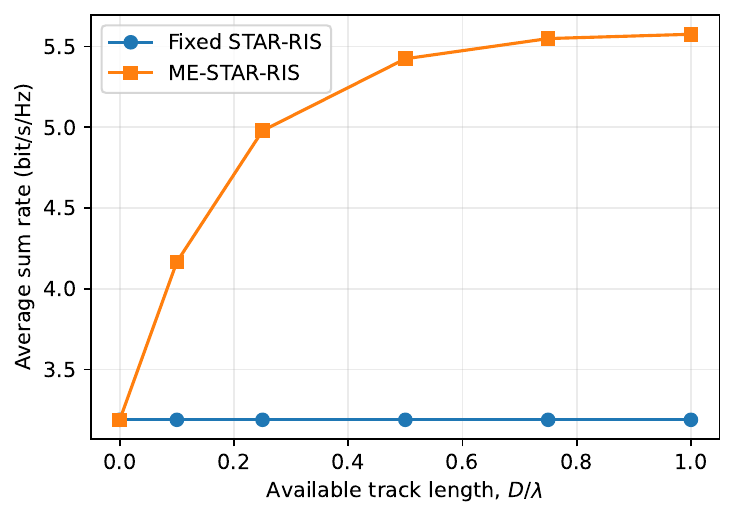}}
\caption{Illustrative 100-realization comparison under identical channel realizations, the same coupled STAR hardware, and the same phase-update budget; ME--STAR--RIS additionally optimizes element positions. Error bars in (a) and (b) denote 95\% confidence intervals. Panel (c) shows sampled non-dominated points from the weighted scalarization and is not claimed to be the globally optimal Pareto frontier. Normalized target illumination power is a one-way probing metric referenced to BS transmit power; its negative dB values reflect propagation loss and it is not a complete two-way radar echo SNR.}
\label{fig:results}
\end{figure*}

\section{Illustrative Case Study: Quantifying the Mobility Gain}
We consider a multi-antenna downlink BS, one communication user on each side of the STAR-RIS, and one transmission-side sensing target. Although Fig.~\ref{fig:architecture} depicts a complete monostatic sensing architecture, the numerical study uses normalized target illumination power, defined as the one-way power delivered toward the target divided by the BS transmit power and reported in dB. Because this normalized power ratio is below unity after the cascaded BS--STAR--RIS--target propagation loss, its dB values are negative; a less negative value indicates stronger target illumination. This metric isolates geometric reconfiguration without introducing radar-cross-section, coherent-processing-gain, or two-way echo-calibration assumptions and should therefore not be interpreted as radar echo SNR.

The fixed and movable surfaces use identical element counts, nominal half-wavelength grids, channel realizations, transmit power, and fast-domain beamforming. Direct BS--user and BS--target paths are assumed blocked so that the comparison isolates the surface-assisted full-space link. ME--STAR--RIS adds only out-of-plane displacement along short tracks normal to the nominal surface; half-wavelength tangential spacing therefore preserves the minimum separation. Position-dependent phases follow the incident and outgoing path-length geometry. Table~\ref{tab:sim} summarizes the compact 28-GHz indoor/roadside model.

For fast control, the BS uses zero-forcing data beams plus a sensing beam in the user-channel null space, with 80\%/20\% data/sensing power. The surface uses equal energy splitting and one feasible branch of the passive coupled response, with reflection phase offset by $+\pi/2$ from transmission. A sensing-aware scalarization combines sum rate with a weighted log-illumination term; the nominal weight is 0.05. For optimization fairness, both architectures share 160 phase-only warm-start updates and then receive the same additional 160-update budget: phase-only refinement for fixed STAR-RIS and joint phase/position refinement for ME--STAR--RIS. Thus, both receive 320 phase-gradient updates under the same local solver. The results are a reproducible local benchmark, not a global-optimality claim.

Figure~\ref{fig:results} reports the 100-realization results under the common phase-update budget described above. In panel (a), mobility improves the rate throughout 20--40~dBm; at 35~dBm with 64 elements, the average sum rate increases from 3.19 to 5.42~bit/s/Hz, or about 70\%. Panel (b) shows that the geometry degree of freedom also improves the mean normalized target illumination power: the advantage is about 0.44~dB at 64 elements and 0.68~dB at 128 elements. These gains arise in a deliberately challenging setting where transmission- and reflection-side users share one coupled STAR response.

Panel (c) sweeps the illumination weight and plots the sampled non-dominated communication--sensing points rather than claiming a global Pareto frontier. At the nominal weight of 0.05, the fixed surface provides 3.19~bit/s/Hz with a mean normalized illumination power of $-123.22$~dB, whereas ME--STAR--RIS provides 5.42~bit/s/Hz with $-122.79$~dB. More importantly, across their overlapping illumination range the movable surface maintains a markedly higher communication rate. Panel (d) reveals a hardware-relevant saturation with available track length $D$: the average rate reaches 4.17, 4.98, and 5.42~bit/s/Hz for $D=0.1\lambda$, $0.25\lambda$, and $0.5\lambda$, respectively, compared with 5.58~bit/s/Hz at $D=1\lambda$. Thus, $D=0.5\lambda$ captures about 94\% of the rate gain available at $D=1\lambda$, while the optimized average absolute displacement is only about $0.13\lambda$.

\section{Practical Challenges and Research Roadmap}
The preceding architecture, control framework, and numerical study illustrate the potential of geometric reconfigurability, but practical deployment requires several hardware-, channel-, and control-level challenges to be resolved. These challenges determine whether the theoretical mobility gain can survive actuation overhead, imperfect environmental knowledge, electromagnetic interactions, and real-time implementation constraints.

\subsection{Movement speed, energy, and reliability}
The first practical question is not ``how far can an element move?'' but ``how often is movement worth paying for?'' Actuation consumes energy, introduces delay, and may reduce hardware lifetime. The correct objective is therefore likely a net ISAC utility that accounts for movement cost, rather than raw rate or sensing performance alone. Subarray/group movement, sparse movable elements, or event-triggered repositioning may capture most of the gain with lower complexity.

\subsection{Channel acquisition after geometry changes}
Moving an element changes the BS--surface channel and the surface--user/target channels. Re-estimating every cascaded coefficient from scratch after every geometry update would negate much of the benefit. Geometry-aware parametric channel models can instead exploit slowly varying path angles, delays, and scatterer positions. Sensing information itself can assist this process: once the network estimates target/user geometry, it can predict channel changes caused by a candidate element displacement.

\subsection{Positioning errors and calibration}
Optimization commonly assumes exact element positions, but mechanical tolerances and sensor errors translate into phase errors, particularly at high carrier frequencies. Robust designs should therefore include bounded or stochastic position uncertainty. Calibration is also required to map commanded coordinates to actual electromagnetic phase response, because the element response may depend on local coupling and substrate conditions.

\subsection{Dynamic mutual coupling}
Element mobility changes inter-element spacing and therefore mutual coupling. This is both a challenge and an opportunity. Ignoring coupling may produce unrealistic optimized geometries with tightly clustered elements. On the other hand, coupling-aware design could deliberately use geometry to obtain a favorable collective response. Electromagnetic-compliant channel/surface models are essential for determining how much of the theoretical mobility gain survives in hardware.

\subsection{Near-field and wideband effects}
At large apertures, spherical-wave propagation and frequency-dependent focusing must be modeled. Position optimization performed at a single center frequency may cause beam squint across a wide bandwidth. Future work should study wideband geometry design that balances multiple frequencies, communication users, and sensing ranges rather than optimizing a narrowband steering vector.

\subsection{Learning-assisted predictive movement}
Movement is most useful when it is predictive rather than reactive. A controller can learn mobility patterns, road geometry, indoor occupancy, or satellite/UAV trajectories and reposition the surface before the channel deteriorates. Machine learning is particularly attractive at the slow timescale, where decisions are infrequent but combinatorial. The fast beamforming layer can remain model-based, providing a hybrid architecture that preserves interpretability and constraint handling.

\subsection{Standardized benchmarking}
Because movable-surface concepts are emerging under several names, benchmark discipline is important. At minimum, studies should report: the physical movement region; element/subarray motion model; minimum spacing; identical footprint and element count for the fixed baseline; transmission/reflection hardware constraints; movement distance or energy; CSI assumptions; and whether near-field/mutual-coupling effects are included. Without these details, gains attributed to ``movability'' can be difficult to interpret.

\section{Conclusion}
ME--STAR--RIS extends the programmable wireless environment from electromagnetic control to joint electromagnetic and geometric control. For ISAC, this additional degree of freedom is attractive because sensing and communication depend not only on transmit power and phase control but also on aperture geometry, channel correlation, focusing, and interference structure. Its value is therefore scenario dependent rather than universal: the largest gains are expected when several spatial objectives compete or when practical hardware and propagation prevent electronic phase control from removing all geometric limitations. The technology is not a replacement for STAR coefficient optimization; it is a slower complementary layer that can place the surface in a more favorable geometric state before fast beamforming is applied. The most credible path forward is therefore a two-timescale, hardware-aware design that quantifies mobility gain against movement cost and uses fixed STAR-RIS as the primary benchmark. With careful modeling of channel acquisition, coupling, positioning error, near-field propagation, and actuation overhead, ME--STAR--RIS can become a useful architecture for full-space 6G sensing and communication rather than merely another optimization degree of freedom.

\balance
\bibliographystyle{IEEEtran}
\bibliography{references}

\end{document}